# A Branched Deep Convolutional Neural Network for Forecasting the Occurrence of Hazes in Paris using Meteorological Maps with Different Characteristic Scales

Chien Wang
Laboratoire d'Aerologie, CNRS and University of Paul Sabatier, Toulouse, France
chien.wang@aero.obs-mip.fr or wangc@mit.edu

**Abstract**
A deep learning platform has been developed to forecast the occurrence of the low visibility events or hazes. It is trained by using multi-decadal daily regional maps of various meteorological and hydrological variables as input features and surface visibility observations as the targets. To better preserve the characteristic spatial information of different input features for training, two branched architectures have recently been developed for the case of Paris hazes. These new architectures have improved the performance of the network, producing reasonable scores in both validation and a blind forecasting evaluation using the data of 2021 and 2022 that have not been used in the training and validation.

## 1. Introduction

The low visibility event (LVE) or haze can interrupt daily life and many economic activities, cause economic loss besides inconvenience for public, and threat human health. The formation of haze in many cases is a result of elevated atmospheric aerosol abundance (*e.g.*, Lee *et al.*, 2017) in combination with favorable weather and hydrological conditions (Wang, 2021). This is because that the scattering and absorption of sunlight by specifically wetted particles can effectively lower the visibility at Earth's surface (*e.g.*, Day and Malm, 2001; Hyslop, 2009).

It is apparent that a good skill in forecasting the occurrence of LVE could effectively mitigate the economic or societal impact of such events. However, this task is still a challenge to many current approaches either based on meteorology or atmospheric chemistry modeling (*e.g.*, Lee *et al.*, 2018; Wang, 2021). In practice, the forecast of low visibility event is typically made for airports often limited for a short lead time (hour or so), commonly using simplified (*e.g.*, one-dimensional, or single column) weather models (*e.g.*, Bergot, *et al.*, 2007), or regression models derived with certain machine learning algorithms and data from a single measurement station (Castillo-Botón *et al.*, 2022). However, significant differences among the predicted outcomes by these simple meteorological models have been identified during model comparison even for the cases under ideal weather conditions, implying issues for generalization of such modeling efforts. On the other hand, results of regression models still display a considerably high absolute error, while in classification, the often adopted 'weighted' f1 scores are only in the mid-70%. Thus, these conventional approaches could easily suffer from the wide variety of weather conditions associated with LVEs that can easily exceed the capacity of a limited data mining effort could offer.

The outcomes from these "traditional" approaches also suggest that we still have insufficient knowledge of the weather or hydrological regimes corresponding to the occurrence of LVEs. As an alternative approach, a forecasting platform based on deep convolutional neural network (CNN), called HazeNet, has been developed and tested on haze events in Singapore (Wang, 2020), and Beijing as well as Shanghai (Wang, 2021). It was trained with a large quantity of meteorological and hydrological maps as inputs and surface visibility data as the targets in a supervised learning. With the high capacity of a deep CNN trained within a high-dimensional input feature space, the established relation between the inputs and LVEs would cover a large variety of different meteorological/hydrological characteristics. In addition, the results could also advance knowledge on the weather and hydrological conditions that favor LVE to occur.

This paper is to report a recent effort to further improve HazeNet and to thus apply its new version to forecast the LVEs in Paris. The paper is organized to present data and training-validation-evaluation methodology after this Introduction. Then, the validation results of the machine will be discussed, followed by the result from an evaluation using 2020 and 2021 data that are entirely independent to the training and validation process. The last section will summarize the effort and major findings of this study.

## 2. Data and Methodology
### 2.1 Data

In a supervised learning procedure, two types of data are needed. The first one is the targeted (often observed) outcomes or labels, either for classification (categorized logic label) or regression (digital data). The second type of data are input features for forecasting the outcomes. Similar to the previous efforts of forecasting low visibility events in Singapore, Beijing and Shanghai (Wang, 2019 & 2021), the surface visibility observations from the site of Charles de Gaulle Airport (CDG) in Paris obtained from the Global Surface Summary Of the Day (GSOD) dataset (Smith *et al.*, 2011) were used to derive labels for the supervised training. While the input features are two-dimensional daily maps of meteorological and hydrological variables over a domain covering a large part of western Europe including France alongside Atlantic Ocean (Fig. 1), obtained from the ERA5 reanalysis data produced by the European Centre for Medium-range Weather Forecasts or ECMWF (Hersbach *et al.*, 2020).

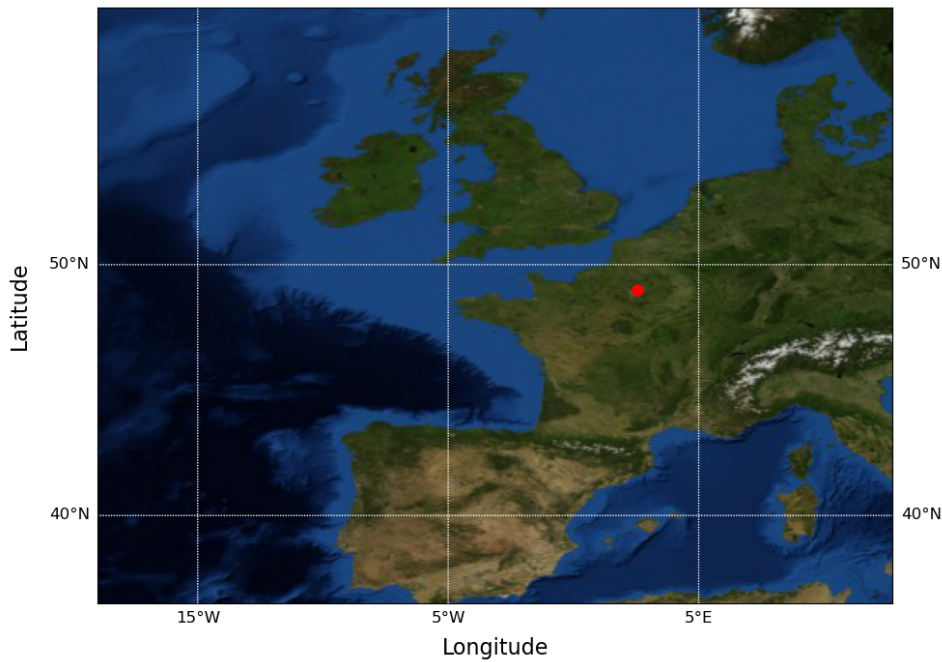

**Figure. 1** The data domain of meteorological and hydrological input features, consisting of 96 latitudinal and 128 longitudinal grids with an increment of 0.25 degree. The red dot marks the location of Charles de Galle Airport.

Comparing to the situations in other metropolitans from Singapore, Beijing, to Shanghai, Paris is rather unique to have similar numbers of 'wet' (fog or mist) and 'dry' (haze without fog, *i.e.*, relative humidity below saturation) LVEs. The most common seasons for LVEs to appear are early spring and winter (Fig. 2). The high-pressure system often occupying France during these seasons brings wet while cold airmass from Atlantic alongside a stable weather condition. A relatively well-developed planetary boundary layer under such a weather condition would benefit the condensation to occur on the surface of aerosols and thus effectively lower the visibility. Such knowledge, however, only provides a general idea for the occurrence of LVEs, as demonstrated in the previous work of Wang (2001), the actual weather conditions associated with the LVEs have a very large variability.

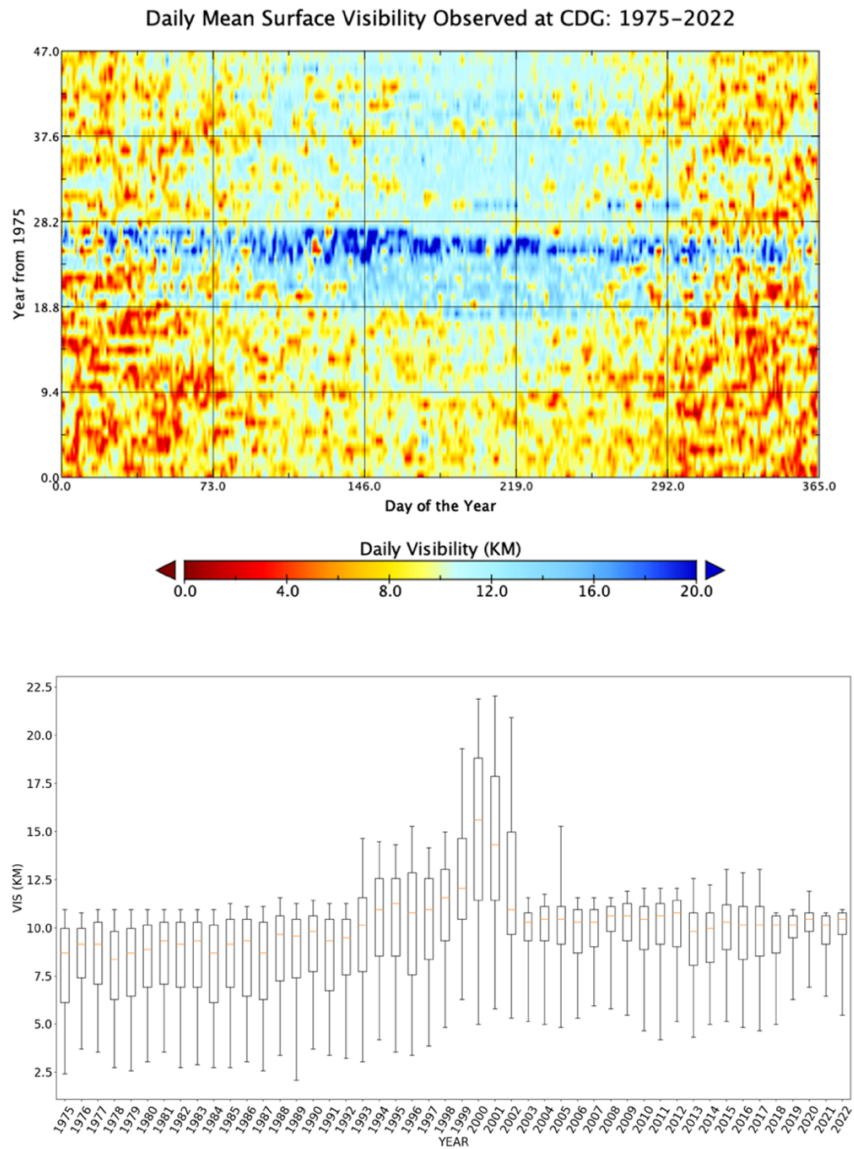

**Figure 2** (a) Daily average surface visibility in km observed at Paris Charles de Gaulle Airport (CDG) since 1975. (b) Whisker plot of annual statistics of daily visibility observed at CDG, the boxes are bunded by the 25$^{th}$ and 75$^{th}$ percentile, extending respectively to the 5$^{th}$ and 95$^{th}$ percentile. The red line marks the annual mean. An unknown systematic switch in statistics during 2000-2002 affects mostly on the clear (high percentile) than haze (low such as the 25$^{th}$ percentile) days.

Using GSOD observations, the labels are derived in both categorized logic format and quantitative format (*i.e.*, visibility in numerical quantity) for different purposes. In the former, LVD is defined as a day with a daily average visibility (*vis.* hereafter) lower than the 25$^{th}$ percentile of the long-term historical observational data (hereafter P25, *i.e.*, 7.89 km for CDG; see Table 1). There were 4,110 such days observed at CDG from 1975 to 2019; among them 1,736 were marked with fog label while 2,374 otherwise, accounted roughly 42% and 58% in the total LVDs, respectively. Because the frequent appearance of fog and misty conditions in Paris, which in many cases are also related to elevated atmospheric aerosol concentrations, all the low visibility days regardless of whether carrying a fog mask in the observation are used to label the LVD events. Note that the GSOD visibility data of CDG contain a systematic shift during 2000 to 2002 (Fig. 2), which affected more on the higher annual percentile than the 25$^{th}$ and 5$^{th}$ percentiles of *vis.* as the latter - the targets of this forecasting effort - only experienced a slightly increase in quantity. This could likely reflect an improvement of the overall air quality, or simply a result of the change of reference targets for the measurement. Therefore, data from these years are kept in the training to maximize the use of available data.

**Table 1.** Certain statistics of the 5$^{th}$, 10$^{th}$, 15$^{th}$, 20$^{th}$, and 25$^{th}$ percentile (denoted as P5, P10, P15, P20, and P25, respectively) in surface daily visibility (*vis.*) of Charles de Gaulle Airport of Paris during 1975 and 2019, 16,436 samples in total.

|  | P5 | P10 | P15 | P20 | P25 |
|---|---|---|---|---|---|
| *vis.* (KM) | 3.70 | 5.15 | 6.44 | 7.24 | 7.89 |
| Events without fog label | 251 | 665 | 1241 | 1788 | 2374 |
| All events | 878 | 1670 | 2579 | 3359 | 4110 |

**Table 2.** Input features, all from ERA5 reanalysis data; here branch 1 is the branch with small kernel and branch 2 with large kernel in HazeNetb2 shown in Fig. 3. Here v18 and v11 represent version with 18 and 11 features, respectively. Paris local time is UTC+1.

| Short Name | Long Name | Branch in v18 | Branch in v11 |
|---|---|---|---|
| REL | Surface relative humidity | 1 | 1 |
| DREL | Diurnal change of REL, [14:19]$_{mean}$ – [6:12]$_{mean}$ | 1 | |
| DT2M | Diurnal change of 2-meter temperature | 1 | |
| T2MS | Daily standard deviation of 2-meter temperature | 1 | |
| U10 | 10-meter wind, zonal component | 1 | 1 |
| V10 | 10-meter wind, meridional component | 1 | 1 |
| TCW | Total column water | 1 | 1 |
| TCV | Total column water vapor | 1 | 1 |
| DTCV | Diurnal change of TCV | 1 | 1 |
| BLH | Height of planetary boundary layer | 1 | |
| DBLH | Diurnal change of BLH | 1 | |
| SW1 | Volumetric soli water, layer 1 | 1 | 1 |
| SW2 | Volumetric soil water, layer 2 | 1 | 1 |
| LCC | Low cloud cover | 1 | 1 |
| Z500 | Geopotential height at 500 hPa level | 2 | 2 |
| D500 | Diurnal change of Z500 | 2 | |
| Z850 | Geopotential height of 850 hPa level | 2 | |
| D850 | Diurnal change of Z850 | 2 | 2 |

The input features are two-dimensional daily fields of selected meteorological and hydrological variables, defined over a 96x128-pixel data domain with a horizontal resolution of 0.25 degree (Fig. 1). The input data have the same number of samples as the labels. They have been normalized in [-1, +1] before use. Initially, 18 variables were selected, which reflect general atmospheric and hydrological status as well as serve as good indicators for weather system, water vapor distributions, atmospheric transport, planetary boundary layer status, and soil moist (Table 2). Then a feature-selection was conducted by screening the style similarity of spatial patterns among these variables. In this procedure, the style losses between every pair of variables throughout the entire sample space have been calculated (see Appendix). When a pair or a group of variables have high style similarity (*i.e.*, low style loss), only one of them is selected to represent that pair or group. As a result, 11 variables out of 18 original ones are selected for the training and validation. The performance of the network using 18 versus 11 variables are very close, with the former has an almost negligible advantage (not shown). Therefore, the results derived using 11 variables will be emphasized in the later discussion unless otherwise indicated.

**2.2 Training-validation-evaluation strategy**

The training and validation of the network use the labels and input data covering the period from 1 January 1975 to 31 December 2019. The entire 16,436 pairs of label-input samples were randomly shuffled first then sampled into two sets, 2/3 of the samples went to the training set and 1/3 to the validation set. As in Wang (2021), these two datasets are reserved only for their designated purpose in practice. Therefore, the validation result is in fact independent to the training process, only serving as an indicator for the performance. The training in this study is relying on a relatively long iteration (typically 2000+ epochs) rather than skills such as early stop or so. This method is to force the performance of the network to converge based on the same validation measure and to minimize overfitting as well.

The trainings in this study have been done in both classification and regression mode. The classification is normally a 2-class training task: class 1 is for the LVD events, defined as a day with *vis*. equal to or lower than the $25^{th}$ percentile (7.89 km; Table 1) of the long-term (1975 to 2019) surface daily visibility observations at CDG, and class 0 for the cases otherwise. For regression, the machine directly predicts a digital value of *vis.* in km. The two different modes of the training also differ in activation of the last dense layer, *i.e.*, sigmoid (2-class; Han and Morag, 1995) or softmax (multi-class; Bridle, 1990) for the classification, linear for the regression. They also use different loss functions, cross entropy (*e.g.*, Hinton and Salakhutdinov, 2006) either binary (2-class) or categorical (multi-class) for the classification, while mean absolute percentage error for the regression. The prediction from regression can certainly be used in performing classification with the same class definition as used in the classification mode training. This also provides a useful comparison between the two different training modes with the same metrics. This latter procedure will be hereafter referred to as regression-classification.

In addition to the training-validation, labels and inputs covering the years of 2021 and 2022, totally 730 samples, obtained from the same data sources as described above, are also prepared for a separate evaluation of the trained network. This evaluation is entirely independent to the training-validation procedure since the machine has never accessed to this dataset before the evaluation. Therefore, it is in fact a blind forecasting test using real-world cases. To distinguish this evaluation from the validation process, the naming of these two procedures will stay the same in the later discussions.

## 2.3 Architecture improvement

The version of HazeNet reported in Wang (2021) is a VGG-type (Simonyan and Zisserman, 2015), 57-layer deep CNN with more than 20 million parameters (Fig. 3, the upper panel). It has been trained and validated using the longitudinal-latitudinal maps of many meteorological and hydrological variables as input features, a dataset with much more channels than the 3-channel RGB images.

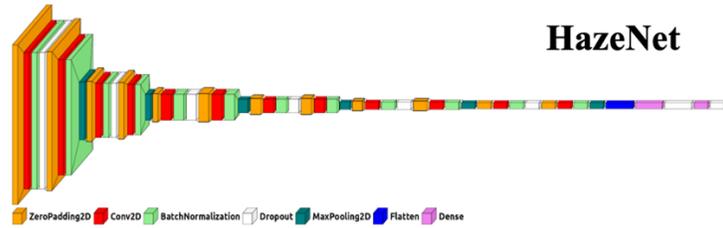

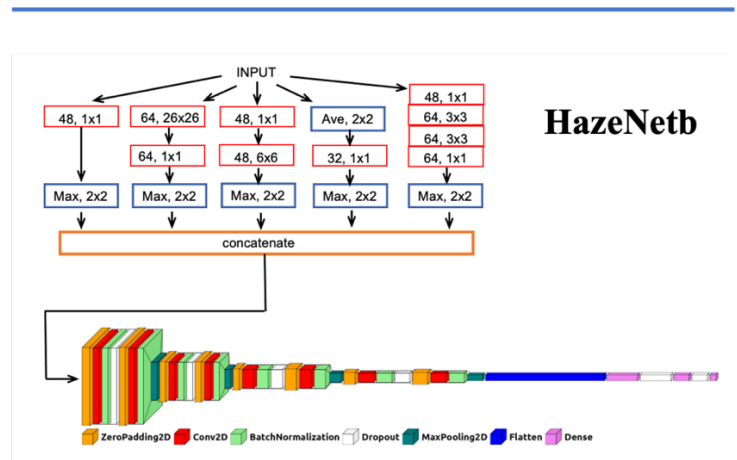

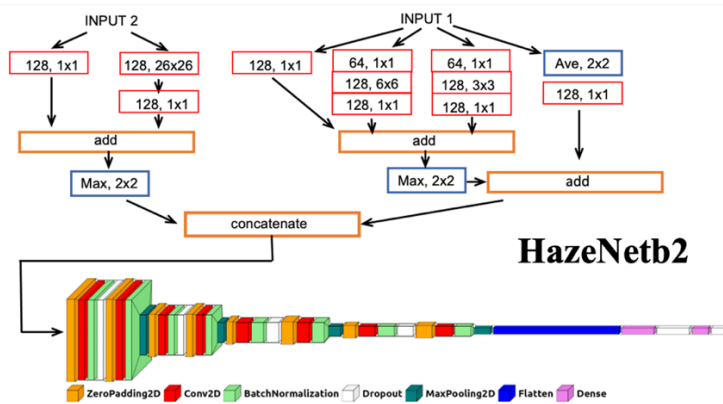

**Figure 3**. Diagrams of different architectures of the networks. Here Max represents a MaxPooling2D layer, Ave represents an Average layer. For 2D convolutional layer, "128, 1x1" represents a such layer using kernel sizes of 1x1 with 128 filter sets. Here each convolutional layer is followed by a batch normalization layer unless otherwise indicated. The bottom parts in HazeNetb2 and HazeNetb are 8-convolutional layers with 3x3 kernels, same as the last 8-convolutional layer of HazeNet. Part of the charts were drawn using visualkeras package (Gavrikov, P., 2020, https://github.com/paulgavrikov/visualkeras).

It is well known that meteorological features have different characteristic spatial and temporal scales (*e.g.*, Holton, 2004; Markowski and Richarson, 2010). For example, the upper panels of Fig. 4 are maps of four meteorological variables of, from left to right: surface relative humidity alongside its diurnal change, 2-meter temperature, and the daily standard deviation of 2-meter temperature. Each of these figures contains many small spatial scale patterns that reflect local weather change. In contrast, four maps in the lower panels, showing geopotential height at 500 (variable 14) and 850 hPa level (variable 16), along with their corresponding diurnal changes (variable 15 and 17), respectively, are consist of smoothly distributed color belts across many grids without showing significant small-scale variations. These features mainly reflect the characteristics of large-scale atmospheric circulation or weather systems, which are less sensitive to the smaller scale factors such as clouds, rainfall, boundary layer waves, and so forth than the features shown in the upper panels. Therefore, a key issue in making CNN to better learn these maps is how to preserve the characteristic scales of different weather or hydrological features in learning.

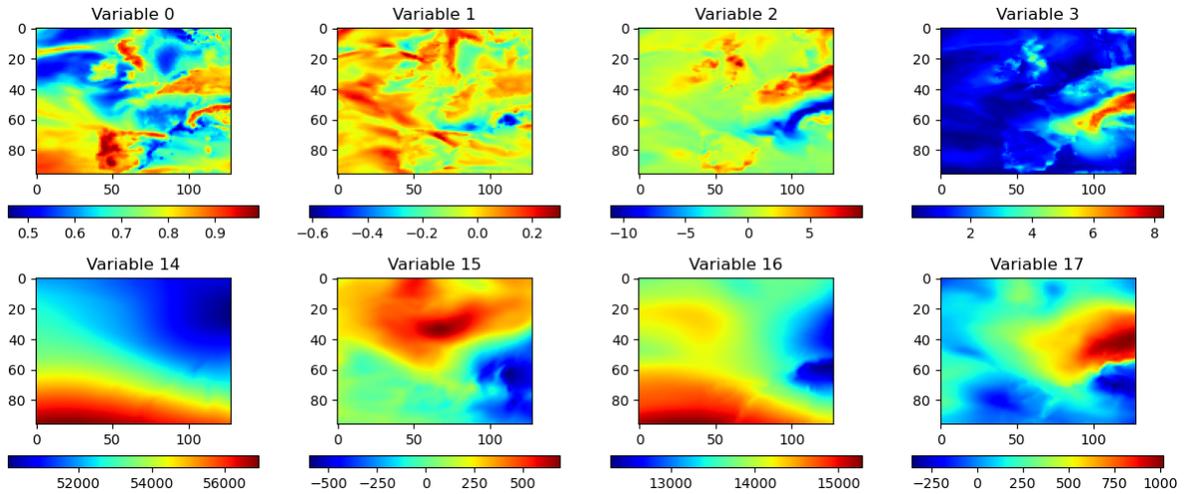

**Figure 4**. Examples of maps of meteorological features with different characteristic spatial scales.

In the previous version of HazeNet, to enable the network to capture patterns with large and medium spatial scales as this might reduce the difficulty in learning, relatively large kernel sizes were adopted for the first two convolutional layers despite the additional computational cost came with it (Wang, 2021). To better preserve the spatial information contained in different features and thus to make the network to better learn various conditions associated with LVEs, a new architecture has been introduced in the HazeNet in this study. The new version of HazeNet is consisted of two parts, the first part is a branched network that processes input features with different kernel sizes, either letting all features go through various branches together (Fig. 1, the mid-panel, *i.e.*, HazeNetb), or letting different features go through different branches based on their characteristic scales (Fig. 1, the lower panel, *i.e.*, HazeNetb2). At the end of the first part of both new architectures, outputs from various branches are concatenated along the filter dimension, then sent to the second part of the network, an 8-layer deep CNN taken from the bottom part of the original HazeNet with 3x3 kernel. The new architectures are apparently inspired by the ideas of Inception Net (Szegedy *et al*., 2015), ResNet (He *et al*., 2015), and 1x1 layer or 'network in network' (Lin *et al*., 2014). Multiple parallel branches with different kernel sizes can lead to an effective learning of pixel-wise patterns or channel wise correlations.

Fig. 5 shows the outputs of the second convolutional layer (just before the 3x3 kernel layer) in the original HazeNet. The displayed patterns contain the color blocks that are likely defined by the features with larger spatial scales, while the distributions of these blocks carry certain influences of the features with smaller spatial scales. On the other hand, from the result of branched HazeNetb (Fig. 6), although the input features contain different characteristic scales, the output of each branch clearly displays a pattern with the characteristic scale resulted from the kernel size used in that branch. Additionally, it can be seen from the outputs of HazeNetb2 in Fig. 7 that such an effect can be further enhanced by arranging different features go through different branches based on their characteristic scales. As demonstrated later, the two new architectures can improve the performance above that of the original HazeNet, while HazeNetb2 is the best performer in most cases among all three architectures.

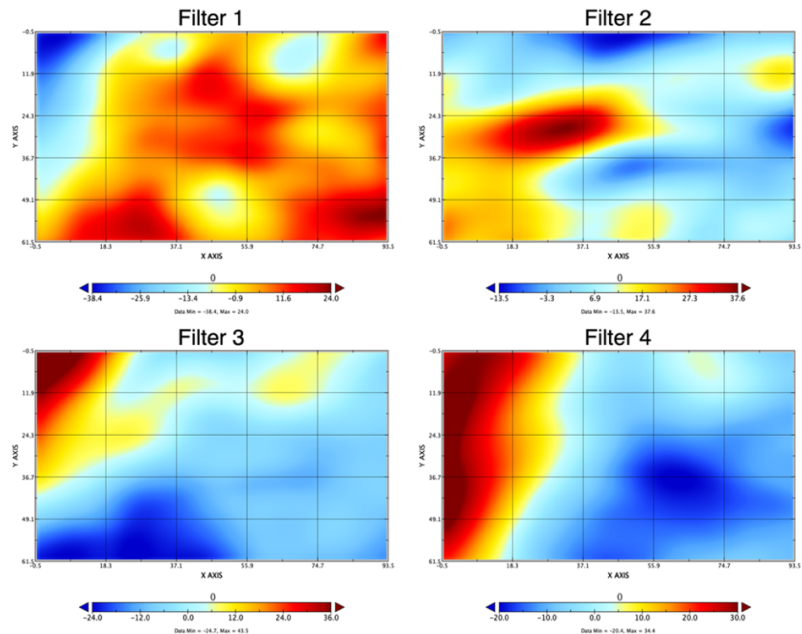

**Figure 5**. The outputs of the first four filters from the second convolutional layer (6x6 kernel) in HazeNet. Note that different color scales are used for various panels to highlight their distributions.

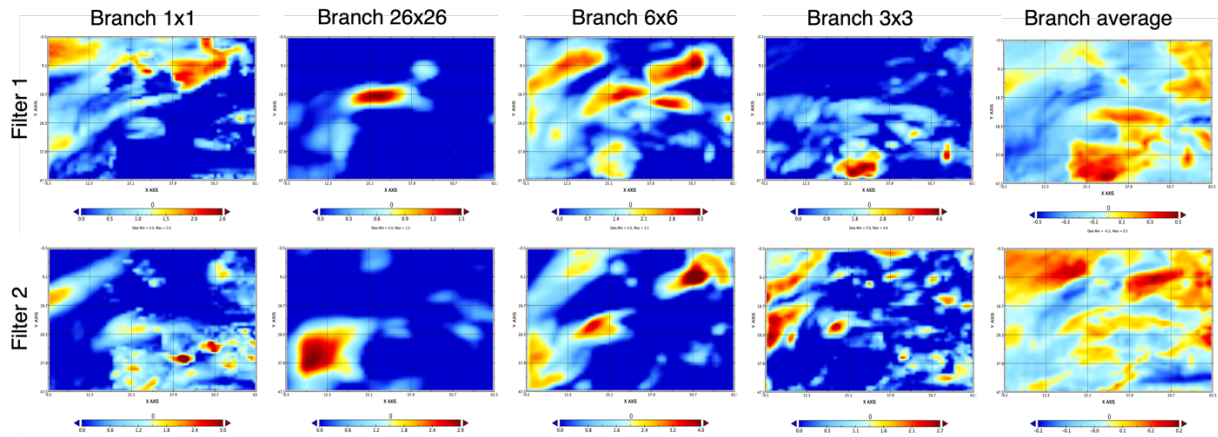

**Figure 6**. The outputs (the first two filter sets) of various branches of HazeNetb just before the concatenate layer (ref. Fig. 3). Different color scales are used for various panels to highlight their distributions.

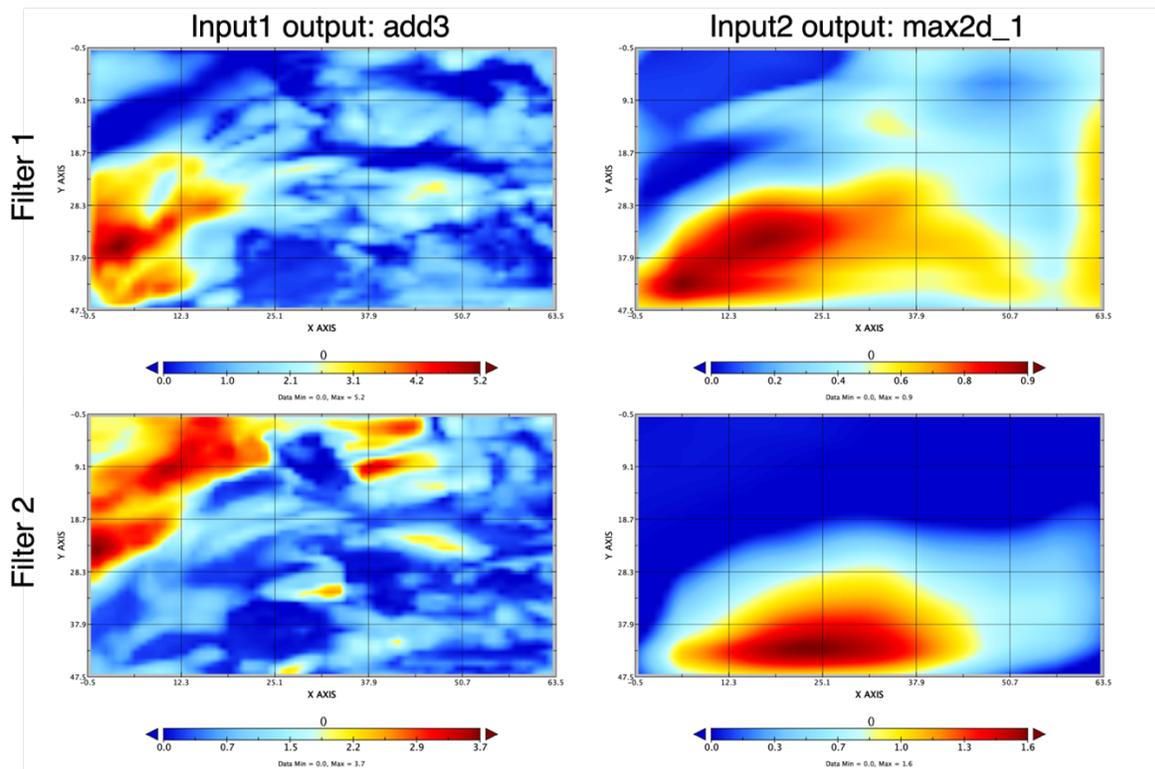

**Figure 7**. (Left) The outputs from the small kernel branch or input1 just before the concatenate layer of HazeNetb2, shown only the first two filter sets. (Right) The same for the outputs from the large kernel or input2 branch. Different color scales are used for various panels to highlight their distributions.

Besides the benefit in preserving patterns with different characteristic scales, the branched architectures can also lower the complexity of the machine. For instance, the number of total parameters of about 20 million in HazeNet has been reduced to near 11 million in HazeNetb, and further to 10.6 million in HazeNetb2. However, due to the utilize of large kernel size, the training time of branched machines are longer than that of HazeNet, with a 10% increase of HazeNetb from that of HazeNet, to a 47% increase of HazeNetb2.

### 3. Performance of Trained Machines

As discussed in the previous section, the length of the supervised training in this study is fixed in 2000 or more epochs that is independent to the validation performance. Using preserved training dataset and validation dataset only for their designated purpose also provides a solid comparison base for all the training tests with different configurations or even architectures.

As demonstrated in Fig. 8, such a training strategy has produced a stable validation performance of the machine. The quantities of all major metrics of classification converge after certain epoch, making a base for fair comparison between different training runs.

The results of different architectures in validation indicate that the branched version HazeNetb2 has delivered the best performance in regression-classification for events with *vis.* lower than the $25^{th}$ or $15^{th}$ percentile of the long-term observations (Fig. 9). While in the regular classification, the difference between the three architectures is rather small. The best performer in above comparison, the newly developed HazeNet2 preserves the characteristic patterns of features with different scales rather well as in Fig. 7, where the outputs of the two branches just before being concatenated clearly show the sharply different patterns with characteristic spatial scales. In addition, regardless of the adopted architecture, the machines in regression-classification mode have produced much better performances comparing to the machines trained in classification mode (Fig. 9 and Table 3).

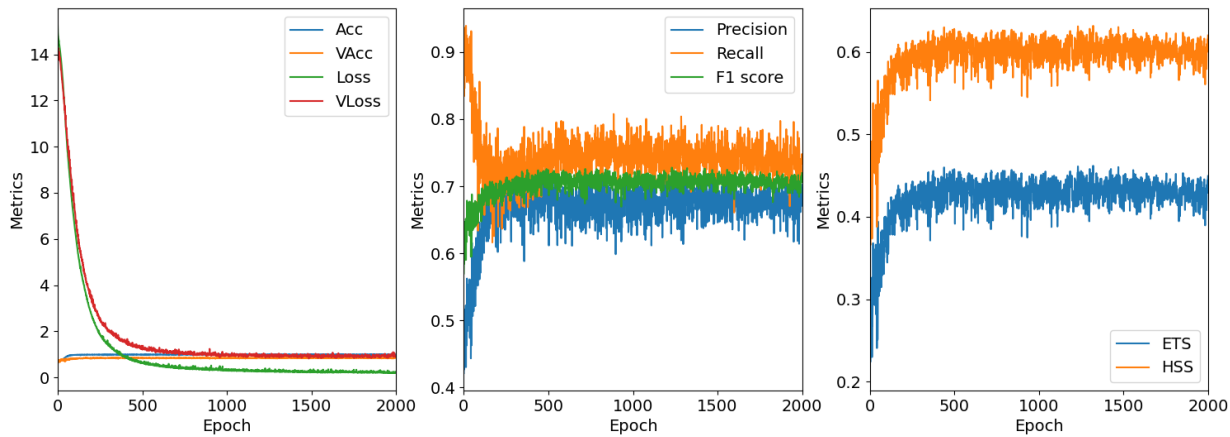

**Figure 8**. An example of the major metrics of performance in validation as a function of epoch. For metrics ref. Wang (2021).

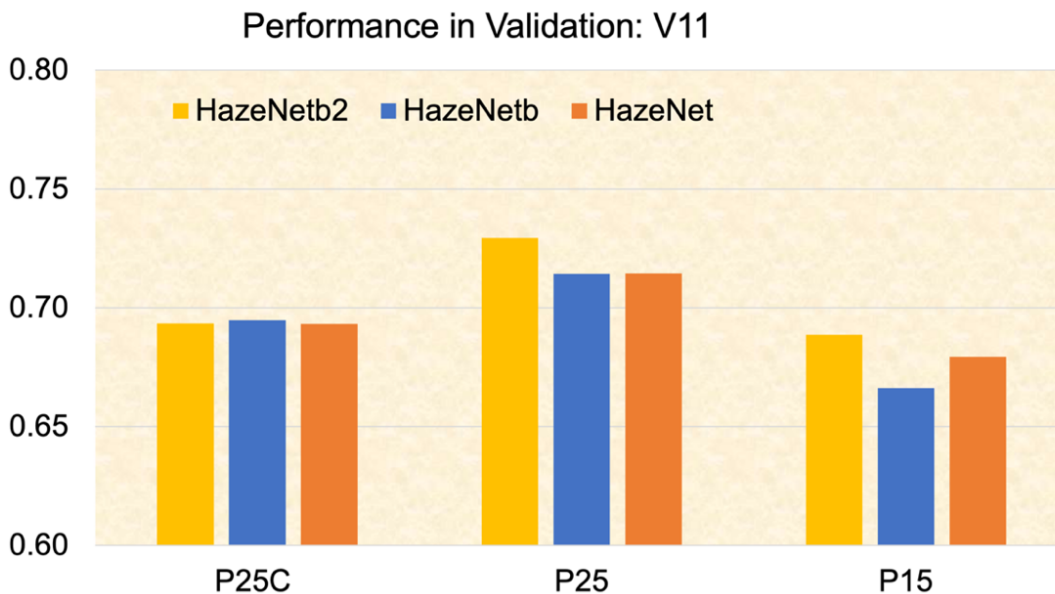

**Figure 9**. The f1 scores of the machines with different architectures. Here P25C represents the results from classification mode, while P25 and P15 the results from regression-classification mode for events lower than either the 25$^{th}$ or 15$^{th}$ percentile of vis. long-term statistics.

**Table 3**. The performance of HazeNetb2 in validation (11 features).

|  | Accuracy | Precision | Recall | F1-Score |
|---|---|---|---|---|
| **Classification (P25C)** | 0.84 | 0.67 | 0.65 | 0.63 |
| **Regression-Classification (P25)** | 0.88 | 0.74 | 0.72 | 0.70 |

## 4. Evaluation of the trained machine using 2021 and 2022 data

To check the performance of the trained machine in making real-world forecast, a "blind forecast" evaluation of various trained then saved machines has been conducted by using the data of year 2021 and 2022 that have not been utilized in either training or validation. During these two years, there were in total 92 LVDs observed at CDG site with a *vis.* lower or equal to P25 or 7.89 km. The evaluation has been applied to check the performance of machines trained in both classification and regression-classification mode.

The evaluation results of machines with different architectures suggest that both branched architectures, *i.e.,* HazeNetb2 and HazeNetb, have achieved a clear improvement comparing to the original HazeNet (Table 4), while HazeNetb2 has delivered the best performance. Again, the regression-classification machines have achieved a more than 10% improvement in f1 score comparing to that of the directly trained classification machine. Regarding the regression machine performance itself, HazeNetb2 scores 8.84% in the mean absolute percentage error, and HazeNetb obtains 9.30%, both clearly lower than the same metrics of HazeNet in 10.59%.

Table 4. The performances of different network architectures in evaluation (11 features).

|  | HazeNetb2 | HazeNetb | HazeNet |
|---|---|---|---|
| **Mean absolute percentage error (%)** | 8.84 | 9.30 | 10.59 |
| $R^2$ | 0.61 | 0.59 | 0.55 |
| **F1 P25C** | 0.67 | 0.65 | 0.63 |
| **F1 P25** | 0.74 | 0.72 | 0.70 |
| **F1 P15** | 0.74 | 0.70 | 0.66 |
| **F1 P5** | 0.67 | 0.65 | 0.60 |

Note: Here $R^2$ = Pearson linear determination coefficient between predicted and observed surface visibility, F1 P25, F1 P15, and F1 P5 represents the f1score in forecasting LVE events with daily *vis.* lower or equal to P25, P15, and P5, respectively, derived from regression-classification, while F1 P25C is the same as F1 P25 except for the direct classification.

As demonstrated in Fig. 10 that the real-world forecasting performance of the trained HazeNetb2 in regression-classification mode are quite impressive. Firstly, the machine has produced an excellent seasonality of LVEs in Paris that matches almost perfectly the observation. Secondly, the point-to-point comparison between predicted and observed *vis.* show a rather small overall discrepancy, or the machine has captured observations reasonably well. For LEVs defined with *vis.* lower than P25 (below the straight black line in the figure), HazeNetb2 in regression-classification mode achieves 0.94 in accuracy, 0.78 in precision, 0.70 in recall, and 0.74 in f1 score (Table 5, P25). For LVEs with *vis.* lower than P15, accuracy, precision, recall, and f1 score is 0.96, 0.61, 0.75, and 0.67, respectively (Table 5, P15). These scores indicate that the machine has captured 70% of the observed LVEs defined by P25 or even 75% by P15, while 78% of the predicted LVEs defined by P25 are true LVEs based on observations. In comparison, trained classification model scores 0.90, 0.57, 0.80, and 0.67 for accuracy, precision, recall, and f1 score, respectively, for LVE events with *vis.* lower than P25 (Table 5, P25C).

Table 5. The performance of HazeNetb2 in evaluation (11 features).

|  | Accuracy | Precision | Recall | F1-Score |
|---|---|---|---|---|
| **Classification P25C** | 0.90 | 0.57 | 0.80 | 0.67 |
| **Regression-Classification P25** | 0.94 | 0.78 | 0.70 | 0.74 |
| **Regression-Classification P15** | 0.96 | 0.61 | 0.75 | 0.67 |

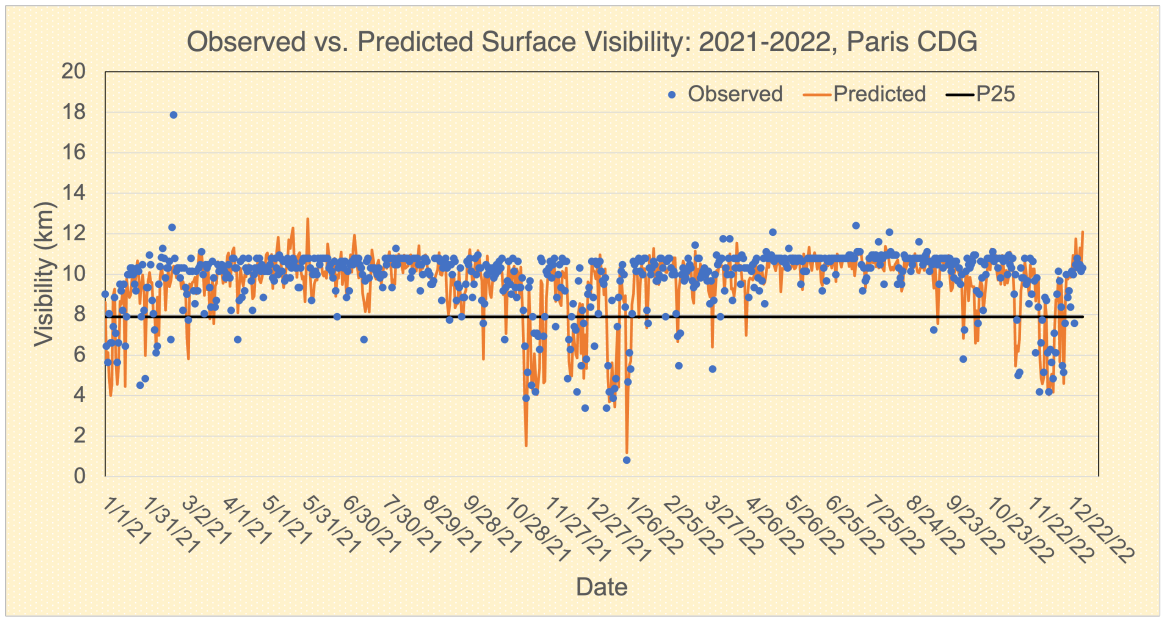

**Figure 10**. Evaluation of HazeNetb2 with 11 features using 2021 and 2022 data.

## 5. Summary and conclusion

A deep learning platform has been developed to forecast the occurrence of the low visibility events or hazes. It uses multi-decadal daily regional maps of multiple meteorological and hydrological variables as input features and surface visibility observations as the targets to train. To better preserve the characteristic spatial information of different features, two branched architectures have recently been developed for the case of Paris hazes.

The branched architectures have lowered the complexity of the machine in terms of the number of total parameters, from more than 20 million to just above 10 million. While the expense of training the machines with branched architectures is clearly increased. Nevertheless, the introduction of the branched architectures has improved the performance of the network in validation and in an evaluation by performing a real-world forecast using data of 2001-2022 that have never been used in training and validation. Specifically in the evaluation, trained machine with branched architecture and using separate path for input features (*i.e*., HazeNetb2) has delivered an impressive performance by scoring 0.74 in f1 score. It has successfully predicted 70% of the observed low visibility events during 2021 and 2022 at Paris Charles de Gaulle Airport, with 78% of predicted low visibility events as the true cases.

# APPENDIX. Calculating the style loss

$$Style\ Loss = \frac{1}{4 \times M^2} \sum_{i}^{M} (S_1 - S_2)^2 \qquad (A1)$$

Here, $S_1$ and $S_2$ is the gram matrix of the three-dimensional (sample, column, row) feature 1 and 2, respectively, $M$ is the size of the features, with gram matrix of feature **f** defined as:

$$gram\ matrix(\boldsymbol{f}) = \boldsymbol{f} \cdot \boldsymbol{f}^T \qquad (A2)$$

**Acknowledgement**.
The work is funded by the l'Agence National de la Recherche (ANR) of France under the Programme d'Investissements d'Avenir (ANR-18-MOPGA-003 EUROACE). The computations of this work were performed using HPC GPU resources of France Grand Équipement National de Calcul Intensif (GENCI), l'Institut du développement et des ressources en informatique scientifique (IDRIS) of CNRS (Grants A0110110966, A0090110966, and A0070110966), and the GPU clusters of French regional computing center CALMIP (Project P18025).

# References

Bergot, T. E. Terradella, J. Cuxart, A. Mira, O. Liechti, M. Mueller, and N. W. Nielsen, 2007: Intercomparison of single column numerical models for the prediction of radiation fog, *J. Appl. Meteor. Clim.*, 46, 504-521.

Bridle, J. S., 1990: Probabilistic interpretation of feedforward classification network outputs, with relationships to statistical pattern recognition, in Neurocomputing: Algorithms, Architectures and Applications (1989; Soulié F.F and Hérault J. eds.), NATO ASI Series (Series F: Computer and Systems Sciences). Vol. 68. Berlin, Heidelberg: Springer. pp. 227–236. doi:10.1007/978-3-642-76153-9_28.

Castillo-Botón, C., D. Casillas-Pérez, C. Casanova-Mateo, S. Ghimire, E. Cerro-Prada, P. A. Gutierrez, R. C. Deo, and S. Salcedo-Sanz, 2022: Machine learning regression and classification methods for fog events prediction, *Atmos. Res.*, 272, 106157. doi:10.1016/j.atmosres.2022.106157.

Day, D. E., and W. C. Malm, 2001: Aerosol light scattering measurements as a function of relative humidity: a comparison between measurements made at three different sites, *Atmos. Environ.*, 35, 5169-5176.

Han, J. and C. Morag, 1995: The influence of the sigmoid function parameters on the speed of backpropagation learning, In *From Natural to Artificial Neural Computation (*Mira, J. and F. Sandoval eds.), Lecture Notes in Computer Science. Vol. 930. pp. 195–201. doi:10.1007/3-540-59497-3_175.

He, K., Zhang, X., Ren, S. and Sun, J., 2015: Deep residual learning for image recognition, arXiv:1512.03385.

Hersbach, H., Bell, B., Berrisford, P., Hirahara, S., Horányi, A., Muñoz-Sabater, J., Nicolas, J., Peubey, C., Radu, R., Schepers, D., Simmons, A., Soci, C., Abdalla, S., Abellan, X., Balsamo, G., Bechtold, P., Biavati, G., Bidlot, J., Bonavita, M., De Chiara, G., Dahlgren, P., Dee, D., Diamantakis, M., Dragani, R., Flemming, J., Forbes, R., Fuentes, M., Geer, A., Haimberger, L., Healy, S., Hogan, R. J., Hólm, E., Janisková, M., Keeley, S., Laloyaux, P., Lopez, P., Lupu, C., Radnoti, G., de Rosnay, P., Rozum, I., Vamborg, F., Villaume, S. and Thépaut, J.-N., 2020: The ERA5 global reanalysis, *Q. J. R. Meteorol. Soc.*, 146, 1999-2049.

Hinton, G. E. and R. R. Salakhutdinov, 2006: Reducing the Dimensionality of Data with Neural Networks, *Science*, 313, 504-507, DOI: 10.1126/science.1127647.

Holton, J. R., 2004: An Introduction to Dynamic Meteorology (4th Ed.), Elsevier Academic Press, Boston, U.S.A. 535pp.

Hyslop, N. P., 2009: Impaired visibility: the air pollution people see, *Atmos. Environ.*, 43, 182-195. doi:10.1016/j.atmosenv.2008.09.067.

Lee, H.-H., Bar-Or, R. and Wang, C., 2017: Biomass Burning Aerosols and the Low Visibility Events in Southeast Asia, *Atmos. Chem. Phys.*, 17, 965-980, doi:10.5194/acp-17-965-2017.

Lee, H.-H., O. Iraqui, Y. Gu, H.-L. S. Yim, A. Chulakadabba. A. Y. M. Tonks, Z. Yang, and C. Wang, 2018: Impacts of air pollutants from fire and non-fire emissions on the regional air quality in Southeast Asia, *Atmos. Chem. Phys.*, 18, 6141–6156, doi:10.5194/acp-18-6141-2018.

Lin, M., Q. Chen, and S. Yan, 2014: Network in network, arXiv, 1312.4400v3 [cs.NE].

Markowski, P., and Y. Richardson, 2010: Mesoscale Meteorology in Midlatitudes, Willey-Blackwell, 430pp.

Simonyan, K. and Zisserman, A., 2015: Very deep convolutional networks for large-scale image recognition, arXiv:1409.1556.

Smith, A., Lott, N. and Vose, R., 2011: The integrated surface database: Recent developments and partnerships, *Bull. Ameri. Meteorol. Soc.*, 92, 704-708, doi:10.1175/2011BAMS3015.1.

Szegedy, C., Vanhoucke, V., Ioffe, S., Shlens, J. and Wojna, Z., 2015: Rethinking the inception architecture for computer vision, arXiv:1512.00567.

Wang, C., 2021: Forecasting and identifying the meteorological and hydrological conditions favoring the occurrence of severe hazes in Beijing and Shanghai using deep learning, *Atmos. Chem. Phys.*, 21, 13149–13166, doi:10.5194/acp-21-13149-2021.

Wang, C., 2020: Exploiting deep learning in forecasting the occurrence of severe haze in Southeast Asia, arXiv:2003.05763 [physics.ao-ph], http://arxiv.org/abs/2003.05763.